\documentclass[lettersize,journal]{IEEEtran}
\usepackage{amsmath,amsfonts}
\usepackage{algorithmic}
\usepackage{algorithm}
\usepackage{array}
\usepackage{textcomp}
\usepackage{stfloats}
\usepackage{url}
\usepackage{verbatim}
\usepackage{graphicx}
\usepackage{cite}
\usepackage{csquotes}
\usepackage{amsmath,amssymb,amsfonts}
\usepackage{xcolor}
\usepackage{booktabs,tabularx}
\usepackage{array}
\usepackage{lineno}
\usepackage{hyperref}
\usepackage{makecell}
\usepackage{subcaption}
\usepackage{caption}
\usepackage{adjustbox}
\usepackage{bm}
\usepackage{multirow}
\usepackage{tikz}
\usepackage{colortbl}
\usepackage{enumitem}
\usepackage[table]{xcolor}

\newcommand{\tick}{\textcolor{green!60!black}{\checkmark}}
\newcommand{\cross}{\textcolor{red}{\(\times\)}}

\begin{document}

\title{Jacobian Voltage Stiffness Metric -- A Measure of Grid-Forming Capability and System Strength in IBR-Dominated Grids}

\author{{Ambuj~Gupta},~\IEEEmembership{Student~Member,~IEEE}, {Balarko~Chaudhuri},~\IEEEmembership{Fellow,~IEEE}, and {Mark~O'Malley},~\IEEEmembership{Fellow,~IEEE}
\thanks{The authors are with the Department of Electrical and Electronic Engineering, Imperial College London, London, UK.}
\thanks{This work was funded by a Leverhulme International Professorship, grant reference [LIP-2020-002] and by the Engineering and Physical Sciences Research Council [EP/Y025946/1].}
}



\maketitle

\begin{abstract}
As power systems transition toward inverter-based resource (IBR)-dominated grids, traditional system strength definitions and metrics are becoming increasingly inadequate to characterize upcoming stability challenges. Emerging definitions characterize system strength in terms of \enquote{voltage source behind impedance (VSBI)} characteristics. Similarly, Grid-ForMing (GFM) IBRs are expected to contribute voltage stiffness by exhibiting near-constant VSBI characteristics in the (sub-)transient time frame. To quantify VSBI characteristics as a measure of system strength or grid-forming capability, this paper proposes the Jacobian Voltage Stiffness Metric (JVSM), derived from the frequency-domain Jacobian. JVSM provides a measure of both small-signal voltage magnitude and phase-angle stiffness. JVSM is demonstrated to serve as a compliance criterion for evaluating the VSBI characteristics of GFM IBRs. When applied for grid strength assessment, it more effectively identifies small-signal stability problems than state-of-the-art strength metrics. The proposed JVSM is validated through electromagnetic transient simulation case studies using the National Laboratory of the Rockies (NLR, formerly NREL) and WECC-approved industry-standard GFM IBR models and on a modified IEEE 39-bus system.   
\end{abstract}


\section{Introduction}
\IEEEPARstart{T}{he} fundamental requirement of power systems is to maintain stable voltage magnitude, phase, and frequency. In the sub-transient time frame, synchronous generators (SGs) inherently behave as ideal voltage sources and exhibit \enquote{voltage-source behind an impedance (VSBI)} characteristics, thereby supporting voltage regulation in conventional power grids. As inverter-based resources (IBRs) replace SGs, the reduction of such voltage-source behavior has raised significant stability concerns. Compared with the commonly used Grid-FoLlowing (GFL) IBRs, Grid-ForMing (GFM) IBRs are seen as a promising solution to address these stability issues.

\subsection{Compliance of VSBI Characteristics of GFM IBRs}
Regulators, system operators (SOs), and industry consortia expect a GFM IBR to provide voltage stiffness to the grid by exhibiting VSBI characteristics in the sub-transient to transient time frame (0--200 ms). This has led to a broadly accepted definition requiring a GFM IBR to behave as a nearly constant internal voltage source behind an effective internal impedance within the (sub-)transient time frame \cite{aemo2023voluntary, ERCOT2024AGS_ESR, miso2024gridforming, pscc_eq_GFM, testing_shahil_shah, ESIG2025_GFM, unifi2026specs}.

For an SG, the flux linkages and associated magnetic-field energy cannot change instantaneously after a disturbance, thus it exhibits ideal VSBI (I-VSBI) characteristics in the (sub-)transient time frame \cite{Machowski2008PowerSystemDynamics}. However, unlike an SG, the VSBI characteristics of a GFM IBR depend on its control design. For example, \cite{pscc_eq_GFM} demonstrates the impact of voltage control bandwidth on the VSBI characteristics of a GFM IBR, showing that a slower voltage control loop can cause the GFM IBR to move away from the expected I-VSBI characteristics. Thus, to assess the VSBI characteristics of a GFM IBR, some SOs mandate and quantify response times as normative requirements \cite{aemo2023voluntary, ERCOT2024AGS_ESR, miso2024gridforming}. The active and reactive power response criteria, such as rise time, peak value, and decay time, are defined in the time domain when the GFM IBR is subjected to a step change in voltage phase, i.e., a phase jump (PJ), or voltage magnitude, i.e., a magnitude jump (MJ), at its point of interconnection (POI). Table~\ref{tab:gfm_requirements} summarizes the detailed time-domain quantitative requirements from three SOs.

\begin{table}[!t]
  \caption{Summary of three system-operator time-domain requirements for GFM IBR VSBI characteristics.}
  \label{tab:gfm_requirements}
  \centering
  \renewcommand{\arraystretch}{1.15}
  \setlength{\tabcolsep}{2.2pt}
  \begin{tabularx}{\linewidth}{lX>{\raggedright\arraybackslash}p{1.3cm}X}
    \toprule
    \textbf{SO} & \textbf{Peak Value} & \textbf{Rise Time} & \textbf{Decay Time} \\
    \midrule
    AEMO \cite{aemo2023voluntary}& 
    $P \geq 0.2$ p.u./$10^{\circ}$ PJ &
    $\leq 15$ ms &
    -- \\
    ERCOT \cite{ERCOT2024AGS_ESR}&
    $\begin{aligned}[t]
      P &\geq 0.2~\text{p.u.}/10^{\circ}\ \text{PJ} \\
      Q &\geq 0.03~\text{p.u.}/3\%\ \text{MJ}
    \end{aligned}$ &
    $\leq 16.6$ ms &
    $\begin{aligned}[t]
      \Delta P &\geq 0 \;\forall t \leq 50~\text{ms} \\
      \Delta Q &\geq 0 \;\forall t \leq 100~\text{ms}
    \end{aligned}$ \\
    MISO \cite{miso2024gridforming}& 
    $P \geq 0.2$ p.u./$10^{\circ}$ PJ&
    $\leq 15$ ms &
    -- \\
    \bottomrule
  \end{tabularx}
\end{table}

However, the current time-domain approaches used by SOs for compliance assessment of GFM IBRs \cite{aemo2023voluntary, ERCOT2024AGS_ESR, miso2024gridforming} have several major drawbacks \cite{gupta2026_IAS}. Since the response times, and thus the time-domain requirements, are at the sub-cycle or few-millisecond level, measurements of active and reactive power, or current, are highly sensitive to the resolution and accuracy of the measurement device, as well as the length of the sliding window used in the moving-average filter. To overcome the drawbacks of time-domain criteria, several studies in the literature have employed frequency-domain techniques to analyze the VSBI characteristics of GFM IBRs \cite{testing_shahil_shah, ESIG2025_GFM, unifi2026specs, avdiaj2025methodology}.


Addressing VSBI characterization in the frequency domain, \cite{testing_shahil_shah} qualitatively defines the expected response of a GFM IBR as nearly constant $P$-$\theta$ and $Q$-$V$ magnitude spectra, with phase close to $180^{\circ}$, over a range from a few hertz (1--5 Hz) to several tens of hertz (20--60 Hz). Extending this, \cite{ESIG2025_GFM, unifi2026specs} propose GFM IBR performance metrics, requiring the positive-sequence impedance $Z_p$ to resemble an R-L branch within $f_1 \pm 40$ Hz, excluding a narrow band around the fundamental frequency, e.g., $f_1 \pm 4$ Hz. They further recommend nearly constant and higher than 0.5 p.u./p.u. $P$-$\theta$ and $Q$-$V$ magnitude within the frequency range of 4--40 Hz. In the same frequency range, the phases should be close to $\pm180^{\circ}$ with a maximum error of $\pm60^{\circ}$. Regarding VSBI characterization, UNIFI \cite{unifi2026specs} also specifies either a passive or low-gain (magnitude $<$ 0.01 p.u./p.u.) for $\omega$-$P$ and $V$-$Q$ characteristics between 4--40 Hz. Similarly, \cite{avdiaj2025methodology} qualitatively compares GFL and GFM VSBI characteristics using the frequency-domain power-flow Jacobian. It recommends testing VSBI characteristics below 50 Hz and suggests $P$-$\theta$ and $Q$-$V$ magnitudes of at least 0 dB (i.e., 1 p.u./p.u.) over the frequency range of interest.

However, based on these works, the requirement to maintain a nearly constant magnitude for the $P$-$\theta$ and $Q$-$V$ spectra can be restrictive. Expecting a minimum $P$-$\theta$ and $Q$-$V$ magnitude at low frequencies (about 1--5 Hz) may exclude GFM IBRs that can exhibit good VSBI characteristics in the sub-transient time frame, thereby limiting the flexibility of GFM IBR control design. The pass/fail compliance criteria elaborated above cannot quantify VSBI characteristics or compare VSBI capabilities across multiple GFM IBRs. Thus, a more suitable assessment criterion and metric are needed to quantify and compare the VSBI characteristics of a GFM IBR. The following subsection further discusses this VSBI characterization for grid-level system-strength assessment.

\subsection{System Strength}

AEMO defines system strength as the ability of the power system to maintain and control the voltage waveform at any bus, both during steady-state operation and following a disturbance \cite{aemo2020systemstrength}. Similarly, NERC defines system strength as the sensitivity of the voltage at a bus to variations in current injections \cite{nerc2017ibrlowshortcircuitstrength}. To quantify system strength, traditional metrics are based on the short-circuit level (SCL), of which the short-circuit ratio (SCR) is the simplest and most widely used. However, SCR and its variants, such as equivalent SCR (ESCR) \cite{fingrid2023escr}, composite SCR (CSCR), and weighted SCR (WSCR) \cite{nerc2017ibrlowshortcircuitstrength}, cannot effectively assess the system strength of IBR-dominated power systems \cite{nerc2017ibrlowshortcircuitstrength, dozein2025systemstrength, huang2026systemstrength}. This is because the underlying assumption in the computation of these metrics is that all GFM IBRs in the power system have VSBI characteristics similar to those of an I-VSBI, such as an SG. However, in IBR-dominated power systems, system strength is determined not only by the power network and SGs, but also by IBRs and their controls. Thus, these traditional system-strength metrics cannot necessarily capture small- or large-signal dynamics in IBR-dominated power systems.

Emerging definitions of system strength categorize it into three broad groups: small-signal, large-signal, and short-circuit strength. VSBI characteristics are becoming a core and increasingly important aspect for assessing all three categories \cite{dozein2025systemstrength, huang2026systemstrength}. Similar to the VSBI requirements for a GFM IBR, VSBI characteristics define system strength as the ability of the grid, at a bus, to reject disturbances and maintain a nearly constant voltage magnitude and phase in the sub-transient to transient time frame. Unlike traditional system-strength metrics, which mainly reflect SCL under fault conditions, VSBI characteristics cover broader power system stability issues, including damping and oscillations, small-signal stability, large-signal stability, and SCL \cite{dozein2025systemstrength}. This is particularly true for IBR-dominated grids, where high fault current is not available to ensure adequate voltage dynamic performance. Thus, there is a need to assess and quantify VSBI characteristics as a system-strength metric.


\subsection{State-of-the-Art Strength Metrics}

Several studies in the literature have proposed metrics to assess the strength of a GFM IBR or the grid in IBR-dominated power systems. Here, some of the relevant metrics are critically reviewed.

\subsubsection{Generalized SCR ($g$SCR)}

The Generalized Short-Circuit Ratio ($g$SCR) is a single-system-strength metric for the entire system that reflects the stability margin of the least damped mode of the system. $g$SCR is defined as the minimum eigenvalue (EgV) of the extended admittance matrix $Y_{\mathrm{eq}}$ \cite{10508461}. However, the calculation of $g$SCR typically assumes that GFL IBRs are current sources and represents all GFM IBRs as ideal voltage sources. As a result, $g$SCR does not capture the detailed IBR dynamics, thereby failing to capture their influence on system strength.



\subsubsection{Impedance Margin Ratio, Admittance Margin}
The Impedance Margin Ratio (IMR) assesses system strength by evaluating the sensitivity of a system mode to variations in the impedance of the IBR connected at that bus \cite{zhu2024impedance}. Using the whole-system admittance/impedance model, IMR captures poorly damped modes and the IBRs with the greatest participation in them. Similarly, Admittance Margin (AM) adapts IMR to assess a system mode's sensitivity to admittance variations at a bus, enabling small-signal strength assessment at buses without existing IBRs \cite{hadjileonidas2024admittance}.

However, obtaining admittance models from frequency-scan data of black-box IBRs via vector fitting (VF) is prone to numerical errors due to model-order selection and artifact poles \cite{shah2022impedancescanwebinar}. Also, in bulk power systems, subsequent extraction of EgVs and residues, and selection of relevant oscillatory modes can be computationally intensive due to high system order, closely spaced modes, and spurious fitted poles.

\subsubsection{Dynamic Impedance, Dynamic SCR, and New-SCR}
Dynamic impedance (DZM) measures GFM IBR strength by assuming a Thevenin equivalent model for the GFM IBR \cite{richwine2023impedance}. DZM is the inverse of the dynamic admittance ($Y_{{adj}}$), which is defined as the phase-adjusted small-signal admittance relating voltage-magnitude perturbations to the reactive-current response of the resource around the 10 Hz frequency band. Extending the concept of DZM to SCR with interaction factors (SCRIF), \cite{bernhardt2026emtmetrics} proposes new-SCR (NSCR) as a measure of system strength. Similarly, \cite{shah2025voltagesourcewebinar} introduces dynamic SCR (dSCR) as the nearly constant value of the $Q(s)/V(s)$ magnitude plot within the range of 4--40 Hz. However, all these metrics assume that GFM IBRs behave as an I-VSBI. Also, since these methods measure strength based on only a single low-frequency value around 10 Hz, they do not capture faster system or GFM IBR dynamics and instead represent only quasi-steady-state dynamics. 

\subsubsection{Grid Strength Impedance Metric, Dynamic Similarity Index, and Forming Index}

The Grid Strength Impedance Metric (GSIM) is defined as a measure of small-signal system strength and is calculated by element-wise multiplication of the eigen-loci of the $2 \times 2$ system admittance matrix ($Y_{\mathrm{sys}}$) and a reference impedance ($Z_{\mathrm{ref}}$) based on a series R-L impedance \cite{henderson2024gsim}. At each frequency, GSIM indicates the system strength relative to the reference impedance. However, the GSIM use case is defined only at the fundamental frequency ($s = 0$), and therefore represents only the steady-state characterization of system strength. Similarly, the Dynamic Similarity Index (DSI) measures how dynamically similar a system is to a chosen reference I-VSBI model \cite{alican2026dsi}. DSI is calculated by subtracting the system admittance from a reference admittance, then taking the maximum singular value (SgV) of the difference over frequency. Both GSIM and DSI require a reference I-VSBI model for their calculation, and only indicate how close the system is relative to that model.

Forming Index (FI) is a GFM IBR strength index that measures how closely a GFM IBR follows or rejects grid-voltage disturbances \cite{zhuang2025quantifyinggridformingbehaviorbridging}. FI is calculated at each frequency by assessing the maximum SgV of the sensitivity of the converter terminal voltage to the grid voltage. However, FI is defined only for a single-GFM IBR infinite-bus setup and depends on the grid-side line inductance and the line dynamic matrix between the converter and the grid. In an IBR-dominated power system, it may not be possible to accurately model the grid as an equivalent Thevenin impedance.

These three metrics reduce the full $2 \times 2$ system admittance matrix to an aggregate measure, potentially obscuring the admittance elements that are responsible for the observed strength. This can be misleading in bulk power systems with high $X/R$ ratios, where voltage-angle and voltage-magnitude stiffness are mainly governed by the off-diagonal terms, $Y_{dq}$ and $Y_{qd}$, or equivalently by the $P(s)/\theta(s)$ and $Q(s)/V(s)$ Jacobian characteristics. For example, two IBRs may show similar GSIM or DSI values even if one differs mainly in the diagonal terms, $Y_{dd}$ and $Y_{qq}$, while the other differs mainly in the off-diagonal terms, $Y_{dq}$ and $Y_{qd}$. Although these cases may appear similar under an aggregate metric, their implications for voltage stiffness and converter-driven stability can differ significantly. As with IMR and AM, these three methods are prone to numerical errors when VF admittance/impedance models from frequency-scan data of black-box IBRs, particularly due to model-order selection and artifact poles \cite{shah2022impedancescanwebinar}. Also, these three metrics are defined at each frequency and do not provide a single system-strength value.

\subsection{Contributions}

To address the need to assess and quantify VSBI characteristics as measures of stand-alone IBR and system strength, this paper proposes the Jacobian Voltage Stiffness Metric (JVSM) as a simple but meaningful proxy of IBR voltage stiffness and system strength. The main contributions are summarized as follows:

\begin{itemize}[nosep,leftmargin=*]

    \item JVSM can be calculated directly from the frequency scans of the black-box IBR or grid models and does not require any modeling information, a reference impedance, a GFL/GFM IBR tag, error-prone VF, or computationally intensive EgV or SgV calculation. JVSM captures the relevant (sub-)transient dynamics without assuming I-VSBI characteristics for the GFM IBR or the grid. Unlike SCR-based and low-frequency-response-based metrics, it accounts for both the speed and magnitude of the power responses to voltage disturbances.
    
    \item JVSM avoids aggregating the system admittance matrix and uses the $P(s)/\theta(s)$ and $Q(s)/V(s)$ Jacobian characteristics to highlight the coupling dynamics important in high $X/R$ bulk power systems.

    \item Unlike existing system-strength metrics, JVSM can be calculated separately for voltage-magnitude and voltage-phase stiffness, while also being combined into a single strength metric. This separation helps characterize stability issues in IBR-dominated power systems, where voltage-magnitude and voltage-phase dynamics can become more granular and less interlinked. This is especially useful for designing and evaluating system services.

    \item JVSM provides a direct way to quantify and compare the voltage-strength contribution that a GFM or GFL IBR provides to the system, enabling device-level GFM IBR compliance assessment. On the system side, JVSM provides a quantitative assessment of grid strength at different locations, enabling grid-strength pre-screening for planning studies and control-room operators.
    
\end{itemize}


\section{Voltage Source Behind an Impedance: Definition and Characterization}

VSBI characteristics describe the ability of a device, or of the grid as seen from a bus, to behave and respond like an I-VSBI at its terminals in the sub-transient to transient time frame. For IBRs, sufficiently large disturbances can trigger current and energy limits; therefore, GFM IBRs are expected to exhibit VSBI characteristics only within their current and energy limits \cite{aemo2023voluntary, ERCOT2024AGS_ESR, miso2024gridforming}. Thus, the expectation of VSBI characteristics from a GFM IBR is limited to small-signal VSBI characteristics. Similarly, from a system-strength perspective, small-signal VSBI characteristics can be understood as the extent to which the voltage magnitude and phase angle are maintained around an operating point following minor disturbances. This work focuses on assessment and quantification of small-signal VSBI characteristics and thus can be used for small-signal stability analysis in IBR-dominated power systems but is not suited to large-signal or fault-level studies. Henceforth in this paper, for brevity, references to VSBI characteristics, response, or strength refer to their small-signal definitions. From a system-strength perspective, the VSBI characteristics, response, or strength of a device are understood to extend to the corresponding characteristics, response, or strength of the grid at a bus.

Grid disturbances can cause abrupt changes in voltage magnitude and phase. To maintain a nearly constant internal voltage behind the impedance, the device should respond naturally and instantaneously to changes in the grid voltage angle and magnitude at its terminals by exchanging appropriate active and reactive power. The speed and magnitude of the response depend on the stiffness of the equivalent voltage source and the equivalent impedance and thus qualitatively characterize the VSBI strength \cite{pscc_eq_GFM}. A device with a rapid and significant active or reactive power response to a voltage PJ or MJ is considered to exhibit strong VSBI characteristics. Conversely, a device exhibiting a slow or low-magnitude response is considered to exhibit weak VSBI characteristics.

In the frequency domain, the active power ($P$) and reactive power ($Q$) responses to small disturbances in voltage phase ($\theta$) and magnitude ($V$) are characterized by the Jacobian matrix \cite{gupta2026_IAS, testing_shahil_shah}. Thus, the frequency-domain Jacobian matrix $J(s)$ can be used to analyze VSBI characteristics and is defined in \eqref{eq:jacobian_power}:

\begin{equation}
\begin{bmatrix}
P(s) \\
Q(s)
\end{bmatrix}
=
\underbrace{
\begin{bmatrix}
\dfrac{\partial P}{\partial V} 
& \dfrac{\partial P}{\partial \theta} \\[6pt]
\dfrac{\partial Q}{\partial V} 
& \dfrac{\partial Q}{\partial \theta}
\end{bmatrix}
}_{\mathbf{J}(s)}
\begin{bmatrix}
V(s) \\
\theta(s)
\end{bmatrix}
\label{eq:jacobian_power}
\end{equation}


In bulk power systems, i.e., transmission grids with a high $X/R$ ratio, the strong coupling between active power and phase angle, and between reactive power and voltage magnitude, is well established. Thus, the VSBI voltage-angle strength depends on the speed and magnitude of the active power response, i.e., the $P(s)/\theta(s)$ Jacobian characteristics in the frequency domain. Similarly, the VSBI voltage-magnitude strength depends on the speed and magnitude of the reactive power response, i.e., the $Q(s)/V(s)$ Jacobian characteristics in the frequency domain. The characteristics of the $P(s)/V(s)$ and $Q(s)/\theta(s)$ become important only for low-$X/R$-ratio distribution grids and are therefore not considered in this analysis, which focuses on assessing system strength in bulk power systems.

The time-domain VSBI characteristics are assessed based on the active and reactive power responses in the sub-transient to transient time frame, i.e., approximately 0--200 milliseconds (ms). Thus, the frequency-domain characteristics need to be analyzed only over the frequency range of approximately 5 Hz and above. Faster network dynamics, characterized by high-frequency components in the frequency domain, have minimal impact on VSBI characteristics and can therefore be ignored in this analysis. At longer time scales, i.e., at lower frequencies, the voltage magnitude and angle are allowed to change to achieve power sharing, reactive power management, and related control objectives. The next section analyzes the VSBI characteristics of an I-VSBI.

\section{VSBI Characteristics of an Ideal Voltage Source Behind an Impedance}

An I-VSBI maintains a constant ideal internal voltage by exchanging the required active and reactive power in response to PJs and MJs at its terminals. SGs behave like an I-VSBI by holding their internal voltage $E^{\prime\prime}$ constant behind an impedance $X^{\prime\prime}$ in the (sub-)transient time frame \cite{Machowski2008PowerSystemDynamics}.
The frequency domain $dq$-domain admittance $Y_{I\text{-}VSBI}(s)$ of an I-VSBI with an impedance $Z = R + j\omega_0 L$, is given in \eqref{eq:Yi_matrix}. For a given operating point, the elements of the Jacobian matrix $J(s)$ are related to those of the admittance matrix $Y(s)$, as shown in \eqref{eq:pq_vm_theta}.

\begin{equation}
Y_{I\text{-}VSBI}(s)
=
\frac{1}{(R + sL)^2 + (\omega_0 L)^2}
\begin{bmatrix}
R + sL & \omega_0 L \\
-\omega_0 L & R + sL
\end{bmatrix}
\label{eq:Yi_matrix}
\end{equation}

\begin{equation}
\begin{bmatrix}
P(s) \\
Q(s)
\end{bmatrix}
=
\begin{bmatrix}
\dfrac{P_0}{V_p} - \dfrac{3}{2} V_p\, Y_{dd}
& - Q_0 - \dfrac{3}{2} V_p^2\, Y_{dq}
\\[6pt]
\dfrac{Q_0}{V_p} + \dfrac{3}{2} V_p\, Y_{qd}
& P_0 + \dfrac{3}{2} V_p^2\, Y_{qq}
\end{bmatrix}
\begin{bmatrix}
V(s) \\
\theta(s)
\end{bmatrix}
\label{eq:pq_vm_theta}
\end{equation}

where $P_0$, $Q_0$, and $V_p$ denote the pre-disturbance steady-state active power, reactive power, and peak terminal phase voltage, respectively. The off-diagonal terms of the Jacobian matrix of an I-VSBI, which represent its active power response to a PJ and reactive power response to an MJ, are derived from \eqref{eq:Yi_matrix} and \eqref{eq:pq_vm_theta} and are given in \eqref{eq:pq_partials}.

\begin{subequations}\label{eq:pq_partials}
\begin{equation}
\left.\frac{P(s)}{\theta(s)}\right|_{V(s)=0}
=
- Q_0
-
\frac{3}{2} V_p^2 \,
\frac{\omega_0 L}{(R + sL)^2 + (\omega_0 L)^2}
\end{equation}

\begin{equation}
\left.\frac{Q(s)}{V(s)}\right|_{\theta(s)=0}
=
\frac{Q_0}{V_p}
-
\frac{3}{2} V_p \,
\frac{\omega_0 L}{(R + sL)^2 + (\omega_0 L)^2}
\end{equation}
\end{subequations}

Given that, under normal operating conditions, IBRs usually operate at a power factor close to unity, the pre-disturbance reactive power is negligible, and the first terms on the right-hand side of \eqref{eq:pq_partials} can be ignored. From \eqref{eq:pq_partials}, it can be observed that, around an operating point, both the $P(s)/\theta(s)$ and $Q(s)/V(s)$ transfer functions of an I-VSBI exhibit standard second-order characteristics, with poles at $s = -\tfrac{R}{L} \pm j\omega_{0}$ and a damped natural frequency of $\omega_d = \omega_0$. For transmission grids with a high $\omega_0 L/R$ ratio, the damping factor ($\zeta$) is low; thus, the resonance frequency satisfies $\omega_r \approx \omega_d = \omega_0$. %
%
%
The steady-state response ($s \to 0$) has a magnitude proportional to the magnitude of the susceptance of the I-VSBI, i.e., $-B = \dfrac{\omega_{0} L}{R^{2} + (\omega_{0} L)^{2}} = \dfrac{X}{Z^2}$. For transmission grids with a high $\omega_0 L/R$ ratio, the steady-state response is proportional to the short-circuit ratio, i.e., $1/(\omega_0 L)$ in p.u..

\begin{figure}[!t]
    \centering
    \includegraphics[width=1\linewidth]{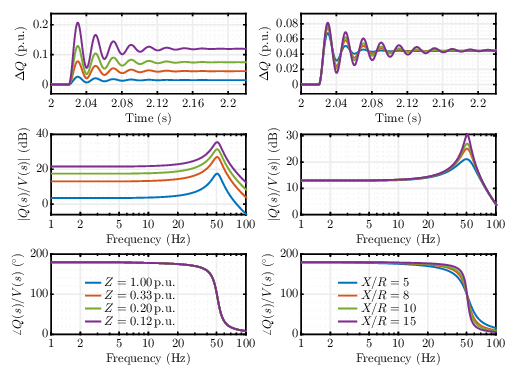}
    \caption{Reactive power step response and $Q(s)/V(s)$ Bode characteristics of an I-VSBI under: (left) varying internal impedance $Z$, and (right) varying $X/R$ ratio.}
    \label{fig:ivsbi_combined_response_bode}
\end{figure}

Fig.~\ref{fig:ivsbi_combined_response_bode} illustrates the change in the reactive power response of an I-VSBI to a 1\% voltage-magnitude step at its terminals and the corresponding frequency-domain $Q(s)/V(s)$ Bode characteristics for: (left) varying the value of the internal impedance $Z$ with an $X/R$ ratio of 10, and (right) varying the $X/R$ ratio with the internal impedance fixed at $Z = 0.33$ p.u.. In both cases, changing the value of $Z$ or the $X/R$ ratio does not affect the speed of response, characterized by time to peak value, with $\omega_d = \omega_0$ remaining constant. 
From the Bode magnitude plots, it can be observed that the resonance frequency, i.e., $f_r = \omega_r/2\pi$, is approximately $\omega_d/2\pi = 50$ Hz. The effect of damping becomes more significant as the $X/R$ ratio decreases, with the reactive power response becoming less oscillatory and exhibiting a lower resonant peak. The Bode phase converges toward $180^\circ$ below the resonant frequency, indicating that the I-VSBI exchanges reactive power to oppose changes in the voltage magnitude at its terminals. At longer time scales, i.e., at lower frequencies, the I-VSBI has no additional control objectives. Thus, the Bode magnitude and phase plots remain constant, and the response has a constant steady-state value. Next, the VSBI characteristics of a GFM IBR are analyzed.

\section{VSBI Characteristics of a GFM IBR}


The GFM IBR control structure can be broadly divided into two control hierarchies: (a) faster dynamics dominated by the inner control loops and (b) subsequent slower dynamics determined by the design of the outer control loops. The outer control loops, including the power plant controller, establish the steady-state relationships among the IBR output active power, reactive power, frequency, and voltage. The faster inner control loops are responsible for limiting the IBR output current and ensuring that its output voltage tracks the references specified by the outer control.
%
%
Since the VSBI requirements are defined within the sub-transient to transient time frame, i.e., 0--200 ms, and the typical control bandwidth of the outer control loops, such as droop control, is approximately 1--5 Hz or slower, the VSBI characteristics of a GFM IBR are largely independent of the outer control loops and depend mainly on the faster inner control loops. However, a faster outer loop characteristic can affect the VSBI strength of the GFM IBR.

\begin{figure}[!t]
      \centering
      \includegraphics[width=\linewidth]{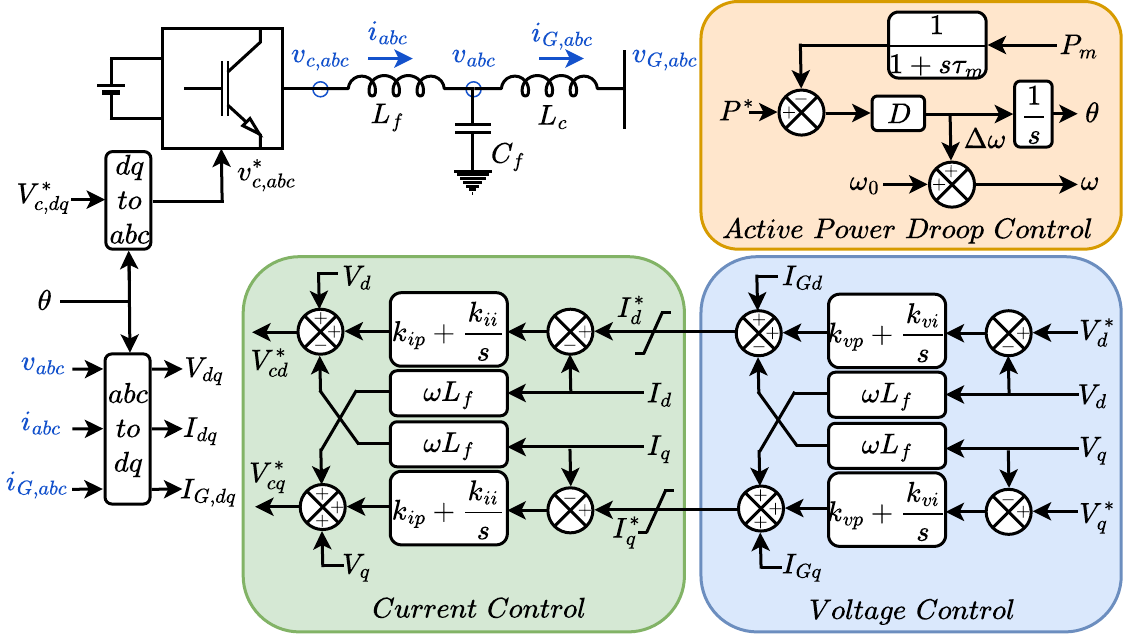}
      \caption{Control structure of the droop-based GFM IBR considered for VSBI analysis.}
      \label{fig:gfm_diag}
\end{figure}

Fig.~\ref{fig:gfm_diag} illustrates a commonly used droop-based GFM IBR model connected to the grid through a coupling impedance. The inner voltage and current control loops use a cascaded proportional--integral (PI) control approach. For simplicity, the dynamics of the primary energy source are omitted from the model, making the study resource-agnostic. To derive the $dq$-domain admittance model and, consequently, the frequency-domain Jacobian matrix for the GFM IBR, the equations representing its small-signal dynamics are described next.

First, assuming ideal decoupling feed-forward compensation \cite{yazdani2010vsc}, the closed-loop current dynamics, including the filter dynamics, are reduced to a first-order lag system with a bandwidth of $1/(2\pi\tau_c)$ Hz, as described in \eqref{eq:current_filter_d} and \eqref{eq:current_filter_q}. For brevity, the small-signal perturbation of each variable, denoted by $\Delta(\cdot)$, is hereafter represented by $(\cdot)$.

\begin{equation}
I_d(s) = G_i(s)\, I_d^*(s) = \left(\frac{1}{1+s\tau_c}\right) I_d^*(s)
\label{eq:current_filter_d}
\end{equation}
\begin{equation}
I_q(s) = G_i(s)\, I_q^*(s) = \left(\frac{1}{1+s\tau_c}\right) I_q^*(s)
\label{eq:current_filter_q}
\end{equation}

The filter-capacitor ($C_f$) dynamics are given in \eqref{eq:cap_dynamics_d} and \eqref{eq:cap_dynamics_q}.

\begin{equation}
I_d = s C_f V_d + I_{Gd} - \omega C_f V_q
\label{eq:cap_dynamics_d}
\end{equation}
\begin{equation}
I_q = s C_f V_q + I_{Gq} + \omega C_f V_d
\label{eq:cap_dynamics_q}
\end{equation}

The combined inner-loop dynamics, from \eqref{eq:current_filter_d} and \eqref{eq:current_filter_q}, are given in \eqref{eq:voltage_filter_d} and \eqref{eq:voltage_filter_q}.

\begin{equation}
I_d = G_i(s)\left[ G_v(s)\left(V_d^* - V_d\right) + I_{Gd} - \omega C_f V_q \right]
\label{eq:voltage_filter_d}
\end{equation}
\begin{equation}
I_q = G_i(s)\left[ G_v(s)\left(V_q^* - V_q\right) + I_{Gq} + \omega C_f V_d \right]
\label{eq:voltage_filter_q}
\end{equation}

where $G_v(s) = \left(k_{vp} + k_{vi}/s\right)$. Assuming constant $V_d^*$ and $V_q^*$ due to the slow outer-loop dynamics, and combining the filter dynamics from \eqref{eq:cap_dynamics_d} and \eqref{eq:cap_dynamics_q} with the inner-loop dynamics from \eqref{eq:voltage_filter_d} and \eqref{eq:voltage_filter_q}, gives the relationship between the grid current ($I_G$) and capacitor voltage ($V$), as described in \eqref{eq:igd_expression} and \eqref{eq:igq_expression}. Here, $\mathcal{A}(s)$ in \eqref{eq:a_expression} is an auxiliary transfer function introduced for notational compactness.

\begin{equation}
I_{Gd} = V_d \mathcal{A}(s) + V_q\left[\omega C_f\right]
\label{eq:igd_expression}
\end{equation}

\begin{equation}
I_{Gq} = V_d\left[-\omega C_f\right] + V_q \mathcal{A}(s)
\label{eq:igq_expression}
\end{equation}

\begin{equation}
\mathcal{A}(s) = \frac{G_i(s)G_v(s) + sC_f}{G_i(s)-1}
\label{eq:a_expression}
\end{equation}

The coupling-inductor ($L_c$) dynamics are given in \eqref{eq:Lc_dynamics_d}, \eqref{eq:Lc_dynamics_q}.

\begin{equation}
V_d = I_{Gd}\left(sL_c\right) - I_{Gq}\left(\omega L_c\right) + V_{Gd}
\label{eq:Lc_dynamics_d}
\end{equation}

\begin{equation}
V_q = I_{Gq}\left(sL_c\right) + I_{Gd}\left(\omega L_c\right) + V_{Gq}
\label{eq:Lc_dynamics_q}
\end{equation}

Combining \eqref{eq:igd_expression}, \eqref{eq:igq_expression}, \eqref{eq:Lc_dynamics_d}, and \eqref{eq:Lc_dynamics_q} gives the simplified relationship between the GFM IBR voltage ($V_G$) and current ($I_G$) at its grid terminals, as shown in \eqref{eq:matrix_relation_gfm_ad}.

\begin{equation}
\underbrace{
\begin{bmatrix}
\alpha & \beta \\
-\beta & \alpha
\end{bmatrix}
}_{\mathcal{M}}
\begin{bmatrix}
I_{Gd} \\
I_{Gq}
\end{bmatrix}
=
\begin{bmatrix}
\mathcal{A} & \omega C_f \\
-\omega C_f & \mathcal{A}
\end{bmatrix}
\begin{bmatrix}
V_{Gd} \\
V_{Gq}
\end{bmatrix}
\label{eq:matrix_relation_gfm_ad}
\end{equation}

where $\alpha = 1 - \mathcal{A}sL_c - \omega^2 C_f L_c$ and $\beta = \mathcal{A}\omega L_c - \omega C_f sL_c$. Simplifying further by taking the inverse of $\mathcal{M}$ gives \eqref{eq:admittance_matrix_GFM}.

\begin{equation}
\begin{bmatrix}
I_{Gd} \\
I_{Gq}
\end{bmatrix}
=
\underbrace{
\frac{1}{\alpha^2+\beta^2}
\begin{bmatrix}
\alpha & -\beta \\
\beta & \alpha
\end{bmatrix}
\begin{bmatrix}
\mathcal{A} & \omega C_f \\
-\omega C_f & \mathcal{A}
\end{bmatrix}
}_{\mathbf{Y^{GFM}}(s)}
\begin{bmatrix}
V_{Gd} \\
V_{Gq}
\end{bmatrix}
\label{eq:admittance_matrix_GFM}
\end{equation}

The small-signal $dq$-domain admittance of the GFM IBR is given in \eqref{eq:Y_GFM_expanded}. The frequency-domain Jacobian of the GFM IBR can then be obtained by combining \eqref{eq:pq_vm_theta} with \eqref{eq:Y_GFM_expanded}.

\begin{equation}
\mathbf{Y}^{GFM}(s)
=
\frac{1}{\alpha^2+\beta^2}
\begin{bmatrix}
\alpha \mathcal{A} + \beta \omega C_f & \alpha \omega C_f - \beta \mathcal{A} \\
\beta \mathcal{A} - \alpha \omega C_f & \alpha \mathcal{A} + \beta \omega C_f
\end{bmatrix}
\label{eq:Y_GFM_expanded}
\end{equation}

The $P(s)/\theta(s)$ and $Q(s)/V(s)$ characteristics, related through $Y_{dq}^{{GFM}}(s)$ and $Y_{qd}^{{GFM}}(s)$, are eighth-order transfer functions. These characteristics of the GFM IBR depend on the speed of the voltage controller ($G_v(s)$), the speed of the current controller ($G_i(s)$), the coupling inductance ($L_c$), and the filter capacitor ($C_f$). Broadly, a larger filter capacitor, a smaller coupling inductor, and faster voltage and current control loops improve the VSBI characteristics of the GFM IBR. The filter capacitor and coupling impedance depend on the GFM IBR size and hardware, and are typically difficult to change. Thus, we illustrate the impact of changing the voltage and current control bandwidths of the GFM IBR.

\begin{figure}[!t]
    \centering
    \includegraphics[width=1\linewidth]{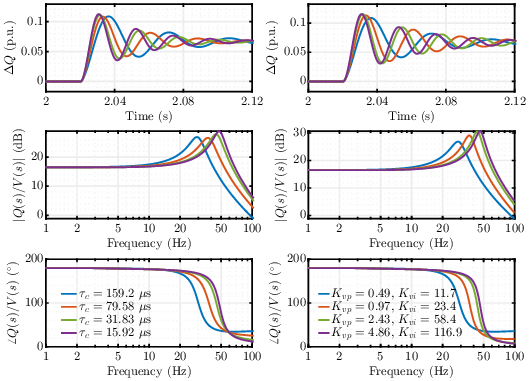}
    \caption{Reactive power step response and $Q(s)/V(s)$ Bode characteristics of the GFM IBR under varying: (left) current control bandwidth, and (right) voltage control bandwidth.}
    \label{fig:gfm_bw_combined_response_bode}
\end{figure}

Fig.~\ref{fig:gfm_bw_combined_response_bode} illustrates the $Q(s)/V(s)$ Bode characteristics for a GFM IBR and the corresponding time-domain reactive power response for changing the: (left) current control bandwidth and (right) voltage control bandwidth for the control design described above.
From the Bode phase plot, a phase of $180^\circ$ is observed below the resonance frequency $f_r$. Increasing the current or voltage control bandwidth shifts the resonance frequency ($f_r$) toward the fundamental frequency ($f_0$), thereby increasing the response speed of the GFM IBR. This can also be confirmed from the corresponding time-to-peak in the time-domain reactive power response. For example, if the GFM IBR current control is tuned to be sufficiently fast, the GFM IBR behaves like an ideal voltage source behind the coupling inductance. Analytically, this can be understood by examining $Y_{dq}$ from \eqref{eq:Y_GFM_expanded} as the current control loop is tuned to be sufficiently fast, i.e.,
$\displaystyle \lim_{G_i(s) \to 1} Y_{dq}(s) = \frac{\omega_0}{L_c(\omega_0^2+s^2)}$,
which is similar to that of an I-VSBI, as described in Section~III.

More generally, depending on its control design, a GFM IBR behaves as a non-ideal or ideal voltage source behind an impedance. The equivalent impedance can change depending on the control design, e.g., due to virtual admittance, the location of the voltage control point, open-loop voltage control, and related factors. However, it is important to note that the Jacobian characteristics can be obtained from a black-box model of the GFM IBR by perturbing the voltage and measuring the power response at its terminals. Thus, assessing VSBI characteristics is independent of knowledge of the GFM IBR control design, and the analysis above for a specific GFM IBR design is used solely to analytically validate its VSBI characteristics. Next, the Jacobian Voltage Stiffness Metric (JVSM) is designed to quantify VSBI strength.

\section{Jacobian Voltage Stiffness Metric (JVSM)}

The voltage stiffness or strength of a device or of the grid can be defined by the speed and magnitude of its active and reactive power responses to changes in the voltage angle and magnitude at its terminals. In the frequency domain, this response is characterized by the $P(s)/\theta(s)$ and $Q(s)/V(s)$ Jacobian characteristics. A device with a rapid and significant active or reactive power response will maintain its internal voltage more constant and is considered to exhibit strong voltage stiffness. Conversely, a device with a slow or low-magnitude response is considered to exhibit weak voltage stiffness. Assuming all devices exhibit I-VSBI characteristics and have a high $X/R$ ratio, SCR, as a traditional index of system strength, captures only the magnitude of the response and assumes a similar speed of response across all devices. This is illustrated in Fig.~\ref{fig:ivsbi_combined_response_bode}, where changing the impedance magnitude, i.e., $Z = 1/\mathrm{SCR}$ or $X/R$ ratio, changes only the magnitude of the response and does not affect the speed of the response. However, as discussed in Section~IV, a GFM IBR can behave as an ideal or non-ideal VSBI and can have varying speed and magnitude of response. Similarly, the grid, as seen from a bus, can exhibit varying responses depending on its topology and the VSBI characteristics of the devices connected to it. Therefore, it is essential to quantify the strength of VSBI characteristics, i.e., to define the voltage stiffness or strength of the grid or a device.

We propose the Jacobian Voltage Stiffness Metric (JVSM) to quantify the system strength of a device or the grid from its frequency-domain $P(s)/\theta(s)$ and $Q(s)/V(s)$ Jacobian characteristics. JVSM is defined as the frequency-weighted area under the $P(s)/\theta(s)$ and $Q(s)/V(s)$ magnitude curves within the frequency range of interest, where the weighting is based on $\log_{10}(f)$. $\mathrm{JVSM}_{V}$ can be calculated as an indicator of voltage-magnitude strength from the $Q(s)/V(s)$ characteristics, as given in \eqref{eq:jvsm-voltage-magnitude}. Similarly, $\mathrm{JVSM}_{\theta}$ can be calculated as an indicator of voltage-phase-angle strength from the $P(s)/\theta(s)$ characteristics, as given in \eqref{eq:jvsm-voltage-phase}.

\begin{equation}
\mathrm{JVSM}_{V} =
\frac{
\displaystyle \int_{f_{\min}}^{f_{\max}}
\left| \frac{Q(s)}{V(s)} \right|
\log_{10}(f)\, df
}{
\displaystyle \int_{f_{\min}}^{f_{\max}}
\log_{10}(f)\, df
}
\label{eq:jvsm-voltage-magnitude}
\end{equation}

\begin{equation}
\mathrm{JVSM}_{\theta} =
\frac{
\displaystyle \int_{f_{\min}}^{f_{\max}}
\left| \frac{P(s)}{\theta(s)} \right|
\log_{10}(f)\, df
}{
\displaystyle \int_{f_{\min}}^{f_{\max}}
\log_{10}(f)\, df
}
\label{eq:jvsm-voltage-phase}
\end{equation}

The $\mathrm{JVSM}_{V}$ and $\mathrm{JVSM}_{\theta}$ metrics can be combined to obtain a single strength metric, $\mathrm{JVSM}$, as given in \eqref{eq:jvsm}.

\begin{equation}
\mathrm{JVSM} =
\sqrt{
\frac{\mathrm{JVSM}_{\theta}^{2} + \mathrm{JVSM}_{V}^{2}}{2}
}
\label{eq:jvsm}
\end{equation}

The calculation of JVSM is done assuming p.u., and thus JVSM itself has units of p.u./p.u. (dimensionless). Regarding the design considerations of the metric, $|P(s)/\theta(s)|$ and $|Q(s)/V(s)|$ capture the magnitude response, while the frequency-weighted integration based on the $\log_{10}(f)$ term emphasizes the response at higher frequencies, i.e., faster response, and thus captures the speed of response. The response is captured in p.u. rather than as Bode gain in dB, so as not to suppress high response magnitudes. The lower bound of the integral, $f_{\min}$, is set to 5 Hz based on the end of (sub-)transient time frame for VSBI characteristics of about 200 ms, and ignores the slower outer-loop dynamics. The upper bound of the integral, $f_{\max}$, is set to 100 Hz to ignore faster dynamics, such as network dynamics. 

Although $\mathrm{JVSM}_{V}$ and $\mathrm{JVSM}_{\theta}$ are defined using only the magnitudes of the corresponding Jacobian frequency responses, the associated phase behavior is largely captured when these characteristics are minimum phase over the frequency range of interest. This is because, for minimum-phase systems, the Bode magnitude and phase are related through the Hilbert transform or Bode gain-phase relationship \cite{ogata2010modern}. From \eqref{eq:pq_vm_theta} and \eqref{eq:pq_partials}, the $P(s)/\theta(s)$ and $Q(s)/V(s)$ characteristics of a device, even an I-VSBI or GFM IBR, may exhibit right-half-plane (RHP) zeros; however, these arise only under weak-grid operation while supplying reactive power for $P(s)/\theta(s)$, or absorbing reactive power for $Q(s)/V(s)$. For typical operating power factors above 0.85 and SCR above 2, these RHP zeros occur at frequencies above several hundred Hz and hence do not affect the Jacobian characteristics in the frequency range of interest. For GFL IBRs, the case studies presented later in the paper demonstrate that the $P(s)/\theta(s)$ and $Q(s)/V(s)$ magnitudes are significantly lower, and thus their contribution to JVSM remains small irrespective of the phase response. When JVSM is evaluated for the grid looking into a bus, the net pre-disturbance reactive-power injection at any scanned bus is zero, which implies minimum-phase behavior of the driving-point $P(s)/\theta(s)$ and $Q(s)/V(s)$ characteristics due to \eqref{eq:pq_vm_theta} and \eqref{eq:pq_partials}. A phase-related penalty, for example, based on group delay or excess phase lag relative to the minimum-phase response, could be incorporated into the metric for theoretical completeness. However, across realistic power-system scenarios, the magnitude-only formulation was found to be adequate (as explained above) and thus provides a simple yet reliable practitioner-oriented pre-screening metric for small-signal stability.



JVSM addresses the gaps in the existing strength metrics detailed in Section I-C, with its advantages over state-of-the-art strength metrics summarized in Table~\ref{tab:stability_metrics}. JVSM provides a single strength value and captures the (sub-)transient dynamics without assuming I-VSBI characteristics for a GFM IBR. Its calculation does not require detailed modeling information, a reference impedance, error-prone VF, or computationally intensive EgV or SgV calculation. Unlike some metrics, JVSM does not mask the important $P(s)/\theta(s)$ and $Q(s)/V(s)$ coupling dynamics in high $X/R$ grids and can be calculated separately for voltage magnitude and phase stiffness.

\begin{table}[!t]
\centering
\caption{Comparison of state-of-the-art strength metrics.}
\label{tab:stability_metrics}
\renewcommand{\arraystretch}{1.08}
\resizebox{\columnwidth}{!}{%
\begin{tabular}{@{}lcccccc@{}}
\hline
\textbf{Criterion} 
& \makecell{gSCR}
& \makecell{IMR\\/ AM}
& \makecell{DZM\,/ NSCR\\/ dSCR} 
& \makecell{GSIM\\/ DSI}
& FI
& JVSM \\
\hline
Single strength value            & \tick  & \cross & \tick  & \cross & \cross & \tick  \\ 
No I-VSBI GFM IBR assumption         & \cross & \tick  & \cross & \tick  & \tick  & \tick  \\
Captures frequency dynamics      & \cross & \tick  & \cross & \tick  & \tick  & \tick  \\
No reference impedance needed    & N/A    & N/A    & \tick  & \cross & \tick  & \tick  \\
Grid-impedance independent       & N/A    & N/A    & \tick  & \tick  & \cross & \tick  \\
No VF, EgV or SgV required         & \cross & \cross & \tick  & \cross & \cross & \tick  \\
Avoids masking coupling dynamics & \tick  & \tick  & \tick  & \cross & \cross & \tick  \\
\hline
\end{tabular}%
}
\end{table}

\section{Application and Case Studies}

In this section, we present two main applications of the proposed JVSM. First, JVSM is used to compare and assess the VSBI-strength compliance of GFM IBRs. Next, JVSM is demonstrated as a grid-strength index for assessing bus-wise grid strength and identifying weak-grid areas at risk of GFL IBR-induced small-signal oscillations.

\subsection{GFM IBR Compliance}

\begin{figure}[!t]
    \centering
    \includegraphics[width=1\linewidth]{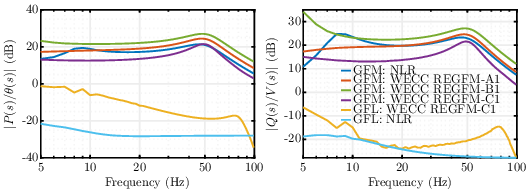}
    \caption{(left) $P(s)/\theta(s)$, and (right) $Q(s)/V(s)$ Bode-magnitude characteristics of industry-standard NLR and WECC-approved GFM and GFL IBR models.}
    \label{fig:pscad_magnitude_comparison}
\end{figure}

To assess the VSBI strength of stand-alone IBRs, we compute and compare the JVSM values of industry-standard EMT-domain GFM and GFL IBR models implemented in PSCAD. These models are taken from the National Laboratory of the Rockies (NLR, formerly NREL) PyPSCAD library \cite{kenyon2021open} and the Western Electricity Coordinating Council (WECC)-approved REGFM model library \cite{pnnl_wecc_gfm_models_2026}. The admittance scan at the terminals of each stand-alone IBR is obtained using the dynamic frequency scanning tool \cite{JIANG2026112226}, and the corresponding $P(s)/\theta(s)$ and $Q(s)/V(s)$ scans are calculated using \eqref{eq:pq_vm_theta}. IBR models are treated as black-box models, and all scans are performed using the default control parameters. Fig.~\ref{fig:pscad_magnitude_comparison} shows the $P(s)/\theta(s)$ and $Q(s)/V(s)$ Bode-magnitude characteristics for each device. It can be observed that, over the frequency range of interest, i.e., 5--100~Hz, all GFM IBRs exhibit significantly higher gains as compared to GFL IBRs.

The $\mathrm{JVSM}_V$ and $\mathrm{JVSM}_\theta$ values for all models are calculated from the respective $Q(s)/V(s)$ and $P(s)/\theta(s)$ Bode-magnitude characteristics and are reported in Table~\ref{tab:jvsm_results}. As expected, a clear separation is observed between the JVSM values of GFL and GFM IBRs, with GFM IBRs exhibiting substantially higher VSBI strength. Since JVSM is a relative strength metric, the values obtained from these industry-standard IBR models can serve as benchmarks for quantifying the minimum expected VSBI strength of GFM IBRs. The SOs can define minimum acceptable values of $\mathrm{JVSM}_V$, $\mathrm{JVSM}_\theta$, and JVSM, according to their system needs, as compliance criteria for VSBI strength, rather than relying on the time-domain approaches whose drawbacks were discussed in Section~I. JVSM provides SOs with a quantitative measure to assess, compare, and incentivize VSBI strength of multiple IBRs from black-box models without relying on GFL/GFM IBR labels.

\begin{table}[!t]
\centering
\caption{JVSM values for industry-standard NLR and WECC-approved GFM and GFL IBR models.}
\label{tab:jvsm_results}
\begin{tabular}{llccc}
\toprule
\textbf{IBR Type} & \textbf{IBR Model} 
& $\boldsymbol{\mathrm{JVSM}_{\theta}}$ 
& $\boldsymbol{\mathrm{JVSM}_{V}}$ 
& $\boldsymbol{\mathrm{JVSM}}$ \\
\midrule
GFM & NLR           & 6.60  & 8.49  & 7.60  \\
GFM & WECC REGFM-A1 & 8.95  & 9.24  & 9.10  \\
GFM & WECC REGFM-B1 & 12.69 & 13.21 & 12.95 \\
GFM & WECC REGFM-C1 & 5.38  & 5.50  & 5.44  \\
\midrule
GFL & WECC REGFM-C1 & 0.15  & 0.11  & 0.13  \\
GFL & NLR           & 0.04  & 0.05  & 0.04  \\
\bottomrule
\end{tabular}
\end{table}

\subsection{Grid Strength Index for Analyzing Small-Signal Stability}

In IBR-dominated power systems, GFL IBR-induced instability in weak grids has become a recognized concern. In weak-grid areas, the GFL IBR terminal voltage is highly sensitive to the injected current and tracks a voltage angle that is itself strongly perturbed by the IBR. This leads to adverse feedback and poorly or even negatively damped oscillations. Thus, weak areas of the grid must be identified and strengthened to avoid GFL IBR-induced instabilities. Here, we demonstrate how SOs could use JVSM as a pre-screening tool to assess grid strength at each location and visualize it using a strength heat map. The grid-strength assessment is validated by examining the spatial distribution of poorly damped modes in the system and identifying the IBRs that contribute to them.

\subsubsection{Single GFL--GFM IBR System}
\begin{figure}[!t]
    \centering
    \includegraphics[width=0.8\linewidth]{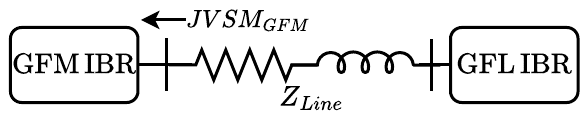}
    \caption{Single GFL--GFM IBR system connected via $Z_{Line}$.}
    \label{fig:SGGS}
\end{figure}

\begin{figure}[!t]
    \centering
    \includegraphics[width=0.9\linewidth]{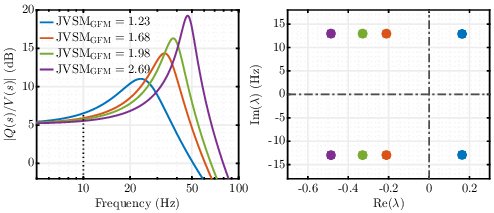}
    \caption{Effect of GFM IBR strength on small-signal stability: (left) $Q(s)/V(s)$ Bode-magnitude characteristics for varying GFM IBR strength, and (right) corresponding eigenvalue loci of the single GFL--GFM IBR system.}
    \label{fig:SGSS_eigen_JVSM}
\end{figure}

Here, we examine the small-signal stability of the single GFL--GFM IBR system shown in Fig.~\ref{fig:SGGS}. In this setup, the GFM IBR control is tuned to vary its strength by changing the voltage and current control bandwidths. As shown in Fig.~\ref{fig:SGSS_eigen_JVSM}(left), this results in different $Q(s)/V(s)$, with the GFM IBR strength quantified by the corresponding $\mathrm{JVSM}_{\mathrm{GFM}}$ values. Fig.~\ref{fig:SGSS_eigen_JVSM}(right) shows the EgVs of the worst-damped modes of the full system. It can be observed that the $\sim$13~Hz sub-synchronous mode is the worst-damped mode and is unstable for the lowest GFM IBR VSBI strength, i.e., $\mathrm{JVSM}_{\mathrm{GFM}}$=1.23. As the GFM IBR strength increases with increasing JVSM, the mode shifts to the left, indicating increased damping and a more stable system. This highlights that the strength of the GFM IBR depends on its control design and that, even with the same network topology, inadequate VSBI strength from GFM IBRs can lead to weak-grid oscillations or instability in GFL IBRs.

From Fig.~\ref{fig:SGSS_eigen_JVSM}(left), it can also be observed that the weakest GFM IBR case, shown in blue with $\mathrm{JVSM}_{\mathrm{GFM}}=1.23$, has a higher $Q(s)/V(s)$, or equivalently $Y_{qd}$, Bode magnitude around 10~Hz. Thus, DZM, or similar NSCR or dSCR metrics, would mischaracterize this GFM IBR as the strongest, which is inconsistent with the EgV analysis.

\subsubsection{IEEE 39-Bus Bulk Power System}

In this subsection, we demonstrate the applicability of JVSM for quantifying bus-wise grid strength in the modified IEEE 39-bus system and for identifying the most vulnerable locations to IBR-induced weak-grid oscillations. The grid JVSM, at a bus, is calculated from the driving-point $P(s)/\theta(s)$ and $Q(s)/V(s)$ characteristics, which are extracted from the driving-point admittance characteristics at that bus for a given operating point. The driving-point admittance characteristics at a bus can be obtained by a dynamic frequency scan at the bus of interest from the whole-system EMT model. Alternatively, the driving-point admittance spectra can be calculated from the available network topology and parameters and the admittance spectra of all devices connected to the grid.

\begin{figure}[!t]
    \centering
    \includegraphics[width=0.9\linewidth]{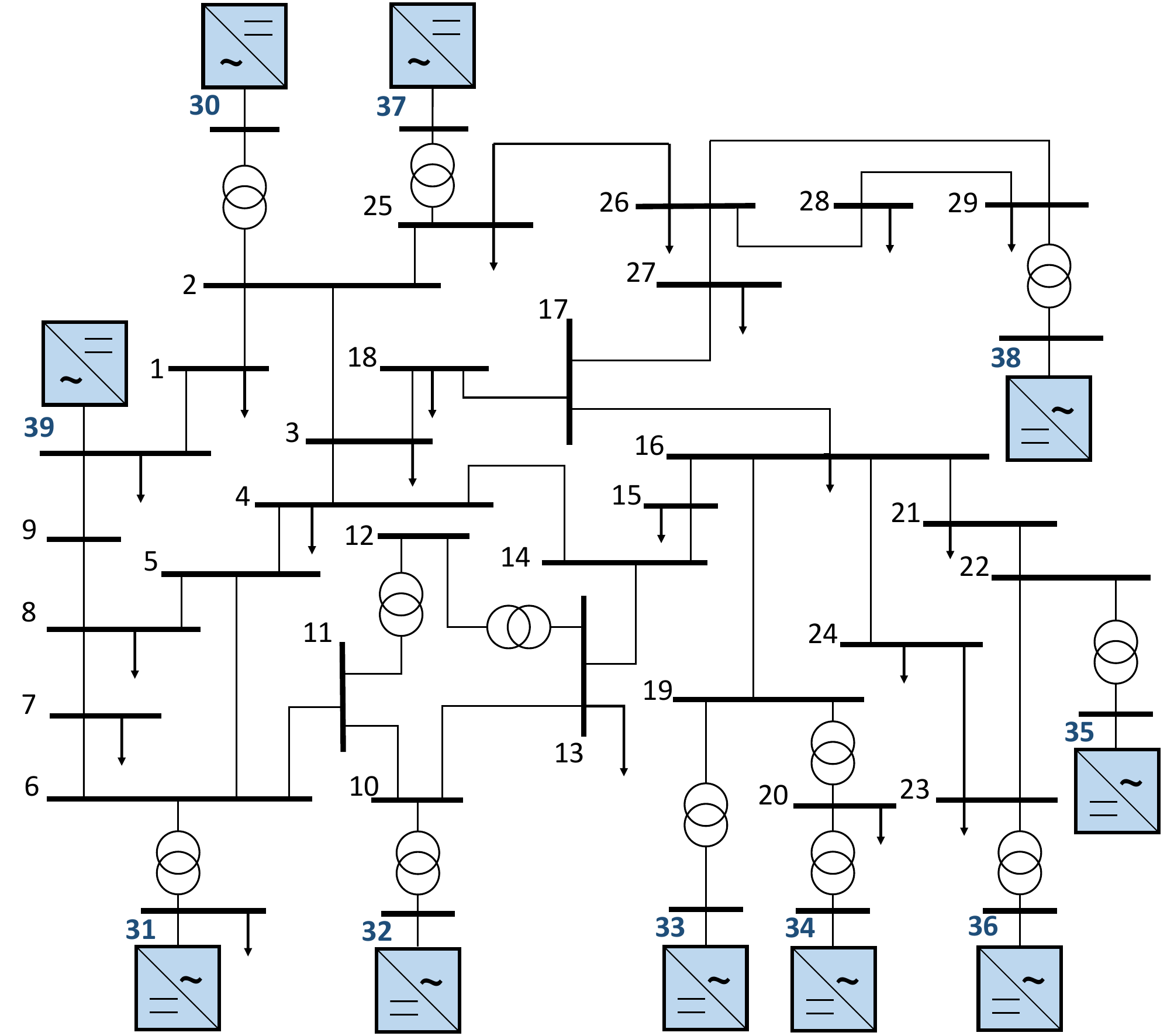}
    \caption{Modified IEEE 39-bus test system with all synchronous generators replaced by IBRs.}
    \label{fig:ieee39}
\end{figure}

The IEEE 39-bus system is modified by replacing all 10 SGs with IBRs, as shown in Fig.~\ref{fig:ieee39}. Based on locational proximity, the IBRs are grouped into four broad zones: buses 30, 37, and 39; buses 31 and 32; buses 33--36; and bus 38. To highlight the impact of VSBI strength on small-signal stability, six IBR-configuration cases are studied, each with two GFM and eight GFL IBRs. These six cases are created by placing the two GFM IBRs in different combinations of the four zones. The IBR configurations for IBR buses 30--39 in the six cases are summarized in Table~\ref{tab:gfl_gfm_scenarios}.

\begin{table}[!t]
\centering
\caption{IBR configurations for the six test cases at generator buses 30--39 of the modified IEEE 39-bus system.}
\label{tab:gfl_gfm_scenarios}
\renewcommand{\arraystretch}{1.15}
\setlength{\tabcolsep}{4pt}
\footnotesize
\newcommand{\zh}{\cellcolor{gray!15}}
\resizebox{\columnwidth}{!}{%
\begin{tabular}{l|ccc|cc|cccc|c}
\toprule
\textbf{Case / Bus} 
& \textbf{30} & \textbf{37} & \textbf{39}
& \textbf{31} & \textbf{32}
& \textbf{33} & \textbf{34} & \textbf{35} & \textbf{36}
& \textbf{38} \\
\midrule
\textbf{Case-A} 
& \zh GFL & \zh \textbf{GFM} & \zh GFL 
& \zh GFL & \zh \textbf{GFM} 
& GFL & GFL & GFL & GFL 
& GFL \\

\textbf{Case-B} 
& \zh GFL & \zh GFL & \zh \textbf{GFM} 
& GFL & GFL 
& \zh GFL & \zh GFL & \zh \textbf{GFM} & \zh GFL 
& GFL \\

\textbf{Case-C} 
& \zh GFL & \zh GFL & \zh \textbf{GFM} 
& GFL & GFL 
& GFL & GFL & GFL & GFL 
& \zh \textbf{GFM} \\

\textbf{Case-D} 
& GFL & GFL & GFL 
& \zh \textbf{GFM} & \zh GFL 
& \zh GFL & \zh GFL & \zh \textbf{GFM} & \zh GFL 
& GFL \\

\textbf{Case-E} 
& GFL & GFL & GFL 
& \zh \textbf{GFM} & \zh GFL 
& GFL & GFL & GFL & GFL 
& \zh \textbf{GFM} \\

\textbf{Case-F} 
& GFL & GFL & GFL 
& GFL & GFL 
& \zh GFL & \zh GFL & \zh GFL & \zh \textbf{GFM} 
& \zh \textbf{GFM} \\
\bottomrule
\end{tabular}%
}
\end{table}

\begin{figure*}[!t]
    \centering
    \includegraphics[width=1\textwidth]{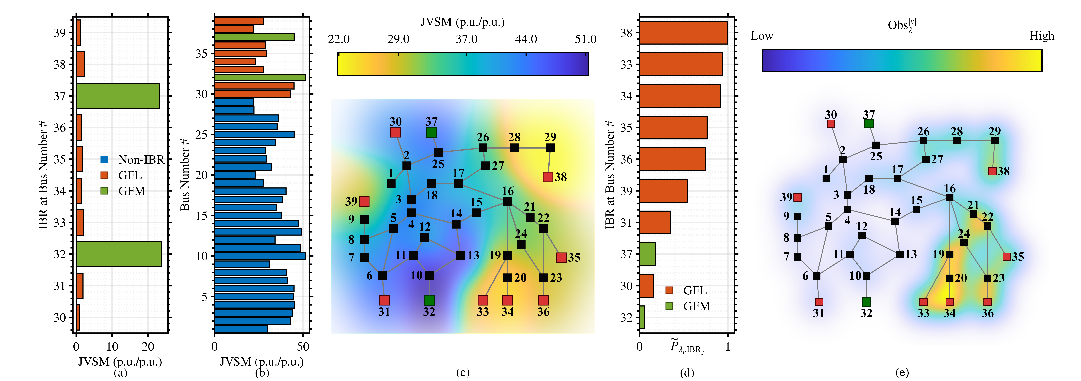}
    \caption{JVSM-based grid-strength assessment and modal validation for Case-A. (a)--(b) JVSM values for stand-alone IBRs and for all grid buses, respectively, showing stronger regions near GFM IBRs and weaker regions in GFL IBR-dominated areas. (c) Spatial heatmap of bus-wise JVSM, highlighting the relative voltage stiffness across the network. GFL and GFM IBRs are highlighted in red and green, respectively. (d) Normalized participation factors of IBRs in the critical 6~Hz oscillatory mode, with GFL IBRs in weak-grid areas showing higher participation. (e) Voltage-magnitude modal-observability heatmap of the critical 6~Hz mode, corroborating that the most vulnerable buses coincide with the weak-grid areas identified by JVSM in (c).}
    \label{fig:case_A_full}
\end{figure*}

For brevity, only the results for Case-A are discussed in detail, while the results of all cases are summarized subsequently. Fig.~\ref{fig:case_A_full}(a) and Fig.~\ref{fig:case_A_full}(b) illustrate the JVSM values for each stand-alone IBR and for each bus in the system, respectively. The stand-alone GFM IBRs have significantly higher JVSM values than the GFL IBRs. The areas around GFM IBRs exhibit higher grid strength, namely buses 6 and 10--14 near the GFM IBR at bus 32, and bus 25 near the GFM IBR at bus 37. This can also be observed from the grid-strength heat map shown in Fig.~\ref{fig:case_A_full}(c). The GFL IBR-dominated areas, namely around buses 33--36 and 38, which are electrically distant from the GFM IBRs, are flagged as weak-grid areas. Thus, the GFL IBRs in these weak-grid areas are expected to be the most probable contributors to oscillations.

To validate the risk of oscillations in weak areas identified from the calculated JVSM, system-level modal analysis is presented next. For Case-A, the EgV analysis flags the 6~Hz mode as the most critical sub-synchronous oscillatory mode, denoted by $\lambda_i$, with a damping ratio of 6.40\%. To analyze the participation of each IBR in this critical mode $\lambda_i$, the system matrix $A^{\mathbf{G}}$ and its associated right and left eigenvectors, $(\psi_i,\phi_i)$, are computed. The entries corresponding to the IBR states are then extracted, and the contribution of $\mathrm{IBR}_j$ to mode $\lambda_i$ is quantified using the summed element-wise participation-factor product over the states of $\mathrm{IBR}_j$, i.e., $|\phi_i^{\mathrm{IBR}_j} \odot \psi_i^{\mathrm{IBR}_j}|$, yielding the participation factor $\mathbf{P}_{\lambda_i}^{\mathrm{IBR}_j}$. The participation factor of each IBR in the 6~Hz mode is then normalized and shown in Fig.~\ref{fig:case_A_full}(d). As expected, the GFM IBRs have the lowest participation in this mode. It can be observed that the GFL IBRs in the weak-grid areas, namely around buses 38 and 33--36, have the highest participation in the critical 6~Hz mode. In contrast, the GFL IBRs in the stronger areas of the grid, namely at buses 30 and 31, have low participation in this mode.

Next, the spatial distribution of the critical 6~Hz oscillatory mode, $\lambda_i$, is examined for Case-A. The modal observability, ${Obs}_{\lambda_i}$, indicates the grid buses and zones that are vulnerable to high-amplitude oscillations. The voltage-magnitude modal observability of mode $\lambda_i$ is calculated as ${Obs}_{\lambda_i}^{|v|}=C^{\mathbf{G}}\psi_i$, where $C^{\mathbf{G}}$ maps the system states to bus-voltage outputs. Fig.~\ref{fig:case_A_full}(e) shows the voltage-magnitude modal-observability heat map of the critical 6~Hz mode at each IBR bus in the system and highlights the most vulnerable areas. It can be observed that the areas around buses 33--36 and 38 are the most vulnerable to the critical 6~Hz mode. These same areas were identified as weak-grid areas from the JVSM heat map. The observability results also corroborate the participation analysis, which shows that the GFL IBRs at buses 33--36 and 38 have the highest participation.

\begin{figure}[!t]
    \centering
    \includegraphics[width=0.9\linewidth]{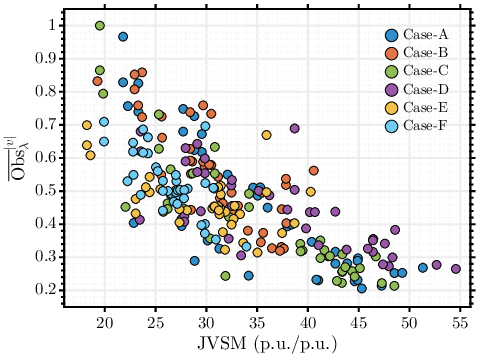}
    \caption{Mean voltage-magnitude observability of the three worst-damped modes versus bus-wise JVSM for the six IBR configuration cases. Higher JVSM values correspond to lower modal observability. The inverse trend demonstrates the use of JVSM as a pre-screening tool for identifying weak buses that are more vulnerable to voltage oscillations.}
    \label{fig:jvsm_mean_observability}
\end{figure}

Next, we validate the results for all Cases A--F with multiple IBR configurations at different locations, as listed in Table~\ref{tab:gfl_gfm_scenarios}. For each case $\chi$, the three worst-damped modes, denoted by $\lambda_{\chi,i}$, $\lambda_{\chi,j}$, and $\lambda_{\chi,k}$, are identified. Then, at each bus $\beta$, the voltage-magnitude observability of these three modes, i.e., $Obs_{\chi,i,\beta}^{|v|}$, $Obs_{\chi,j,\beta}^{|v|}$, and $Obs_{\chi,k,\beta}^{|v|}$, is calculated. The mean voltage-magnitude observability at bus $\beta$ for case $\chi$, denoted by $\overline{Obs}_{\chi,\beta}^{|v|}$, is calculated as $(Obs_{\chi,i,\beta}^{|v|}+Obs_{\chi,j,\beta}^{|v|}+Obs_{\chi,k,\beta}^{|v|})/3$. For each case and each bus, $\mathrm{JVSM}_{\chi,\beta}$ is compared with $\overline{Obs}_{\chi,\beta}^{|v|}$ to assess the vulnerability of weak and strong buses to voltage oscillations. Fig.~\ref{fig:jvsm_mean_observability} shows the corresponding scatter plot. It can be observed that stronger buses, quantified by higher JVSM values on the x-axis, are less vulnerable to voltage oscillations. In contrast, weaker buses with lower JVSM values exhibit higher observability and are therefore more vulnerable to voltage oscillations. This validates the use of JVSM as a pre-screening tool for quantifying grid strength and identifying areas vulnerable to poorly damped sub-synchronous oscillations. 

%

\section{Conclusion}

This paper shows the effectiveness of the Jacobian Voltage Stiffness Metric (JVSM) as a relative measure for quantifying small-signal voltage-source-behind-impedance (VSBI) characteristics in IBR-dominated power systems. JVSM provides a unified measure for quantifying both device-level VSBI strength and system-level voltage stiffness in IBR-dominated grids. Unlike existing strength metrics, JVSM does not assume I-VSBI behavior, require a reference impedance, or a detailed model. JVSM can be computed directly from black-box frequency scans, eliminating the need for error-prone vector fitting and computationally intensive eigenvalue or singular value analysis. The demonstrated applications of JVSM include device-level compliance assessment and system-level grid-strength screening. At the device level, JVSM provides a clear quantitative distinction between GFM and GFL IBR behavior, indicating that it can serve as a practical compliance-oriented metric for comparing the VSBI strength of IBRs. At the system level, JVSM can be used as a grid-strength pre-screening tool to provide an intuitive and computationally efficient indication of locations that are more vulnerable to converter-driven oscillations. JVSM heat maps can identify weak regions in IBR-dominated grids that are electrically distant from devices with voltage source behavior and may exhibit high modal observability of the worst-damped modes. Overall, system operators can use JVSM to support GFM IBR compliance assessment, strength-service specification, and weak-area identification for planning and operational studies.





\bibliography{JVSM_References}

\end{document}